# Running Genetic Algorithms on Hadoop for Solving High Dimensional Optimization Problems

Güngör YILDIRIM, İbrahim R. HALLAC, Galip AYDIN, Yetkin TATAR
Department of Computer Engineering
Firat University
Elazig, Turkey

*Abstract*—Hadoop is a popular MapReduce framework for developing parallel applications in distributed environments. Several advantages of MapReduce such as programming ease and ability to use commodity hardware make the applicability of soft computing methods for parallel and distributed systems easier than before. In this paper, we present the results of an experimental study on running soft computing algorithms using Hadoop. This study shows how a simple genetic algorithm running on Hadoop can be used to produce solutions for high dimensional optimization problems. In addition, a simple but effective technique, which did not need MapReduce chains, has been proposed.

*Index Terms*—**Hadoop, MapReduce, Genetic Algorithm**

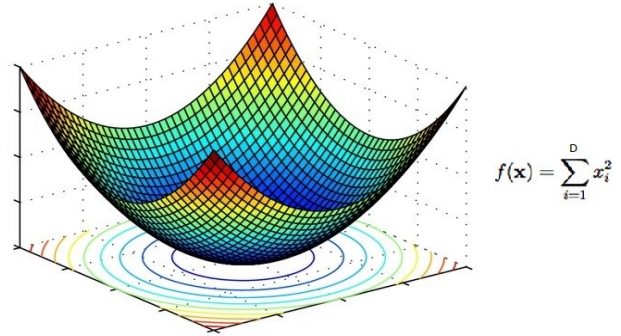

Fig. 1. The Sphere Function

## I. INTRODUCTION

Hadoop gives a new impulse to parallel and distributed systems, which are one of the most interesting study areas of information technologies [1,2]. It provides an easy to use programming API, which can be used to create simple solutions for many complex problems. Hadoop uses MapReduce (MR) programming model, which is the popular Google approach for big data analysis [2,3]. Hadoop can also be used to apply soft computing methods on big data. The so-called Hadoop Ecosystem also provides a comprehensive machine-learning library called Mahout [4].

In this paper we describe our approach to solve a high-dimensional optimization problem by employing a soft computing method using Hadoop. In short, a simple genetic algorithm (GA) which used very big populations was programmed using Hadoop MapReduce API, and applied to a high-dimensional optimization problem. We first surveyed the literature for different MapReduce-Genetic Algorithm (MRGA) models and proposed a simple but effective approach. In the experiments, the sphere function, which is a well-known benchmark function in the literature was selected (Fig. I).

To date, many different approaches regarding soft computing methods have been proposed for various high-dimensional problems [5-8]. Those are either novel or hybrid methods, and are generally performed on a single PC. In this context, parallel and distributed systems can be utilized as an alternative for high-dimensional problems.

## II. RELATED WORK

Although Mahout provides a MapReduce based GA implementation, several different MRGA models also exist in the literature [4-8].

Among them, two types of approaches generally stand out. The first type performs the calculations in the Map phase, the second uses Map-Map chains or Map-Reduce chains. The first approach is quick because it produces the population and run GA completely on the memory, but the population size is limited with the size assigned to a map. The second can produce a big HDFS (Hadoop Distributed File System) [5-6] output, which causes a performance loss, but is more flexible and enable usage of different components like combiner.

The model used in this study (Fig.2) can be considered as a hybrid version of the two aforementioned models. But here, first populations were produced on HDFS in order to evaluate the Read-Write performance and so that the populations could be reused in different experiments. The GA process firstly performed in Map phases. In the first experiments, we tried a standard MapReduce process for each population. But, as can be seen in the next section, successful values were not obtained. Then, we suggested that more elite individuals selected from each final population in the Maps by a very low selection rate could be passed on a new population in the reduce phase. Thus a small mixed-eligible population was created. Since the size of the elite individuals passed on was small, performance loss was not too high in comparison with the chain model.

The main population, P, is divided into blocks on HDFS and Maps uses them as a Map population, Mp(i). The pseudo code and model are given in Fig II. Here, C, MCp, RCp represent chromosome of the first population and chromosome of the Map populations, respectively. *F* is the fitness function and, *I* represents the generation number.

---

For $a \in \mathbb{R}$ $C=[a_0, a_1,...a_D]$ and $C \in P$
$MCp \in Mp$ and $Mp_{(i)} \subseteq P$
$RCp \subseteq Mp_{(1)} \cup Mp_{(2)} \cup ... \cup Mp_{(i)}$
$N=[1,2...I]$

---

*Mapper*

**UCp, evaluated and updated Mp population**

**Map (key,velue,context)**
  *While MCp available in Mp*
    *UCp=Evaluate(Update(get(MCp)))*
    *Add(Mp, UCp)*
  *end while*

*function cleanup (context) of Map*
    *while N<I*
      *Rank(Cp)*
      *Crossover(), Mutation(),*
  *Evaluate()*
      *N++*
    *Endwhile*
  *S=% elite selection rate*
  *Emit(key_eliteCp, value_eliteCp) of S%*

---

*Reducer*

**RUCp, evaluated and updated Mp population**

**Reduce(key,velues,context)**

  *While elite chrm. available*
    *RUCp= get(key_eliteCp,value_eliteCp)*
    *Add(Mp, UCp)*

  *end while*

*function cleanup (context) of Map*
    *while N<I*
      *Rank(Cp)*
      *Crossover(), Mutation(), Evaluate()*
      *N++*
    *Endwhile*
  *Emit(the result)*

Fig. 2. Pseudo-Code of the MRGA

## III. HADOOP

In this study we used the open source Apache Hadoop framework, which is developed in accordance with the information given in the [3] paper where the MapReduce model was initially explained. MapReduce is a programming model developed by Google for analyzing big data on distributed computer clusters. In this model a two-phased action called map and reduce is performed on data. Mappers list the data as key-value pairs, which is distributed in blocks on the distributed files system. Let k1 be input key, v1 be input value, k2 be output key, v2 be intermediate value and v3 be output value:

$$\text{Map }(k1,v1) \rightarrow \text{list }(k2,v2)$$

After intermediate key/value lists are produced, Reducers perform grouping operation on key/value pairs.

$$\text{Reduce }(k2, \text{list }(v2)) \rightarrow \text{list }(v3).$$

In order to run a program written in MapReduce programming model, there should be an execution environment containing the classes, which are written, based on this model. In study we used Hadoop version 2.6.0 and it should be considered that the information regarding the Hadoop framework given here might not apply to other versions.

Another important component for running a MapReduce program is a distributed file system. Computers, which can communicate with each other via a network connection, are needed to establish a distributed file system, which is necessary for distributing the data on to the machines, processing and writing the results. The file system of Hadoop, which is called Hadoop Distributed File System (HDFS) [9,10], was developed based on Google File System (GFS). HDFS allows us to work with big data on distributed clusters.

In Hadoop system data is split into blocks. The size of each block is set to 128 MB (64 MB for older version than 2.2.0) by default [16]. Different block sizes can be configured using dfs.block.size parameter. These blocks are replicated and then distributed on the computers in the Hadoop cluster. Replication factor can be specified by changing the parameters in the hdfs-site.xml file. Thus, each data has a backup in block level and when there is failure on a machine where some part of the data is stored; the program execution is not affected and data loss doesn't occur [9,10]. In our work we set the replication number as 2.

In this study Hadoop cluster was deployed on a private cloud. The instances of the clusters are virtual machines, which are created using the OpenStack Cloud Computing Software [11,12]. OpenStack runs on a server with 1 TB of hard disk, 50 GB memory and 4 CPU cores. 5 virtual machines were created on OpenStack Cloud. These machines are named as master, slave1, slave2, slave3 and slave4. A general overview of this deployment is shown in Fig. 3. The master machine has 8 GB of memory and each slave machine has 4 GB of memory. In a Hadoop cluster master machine functions as NameNode and JobTracker and slave machines function as DataNode and TaskTracker.

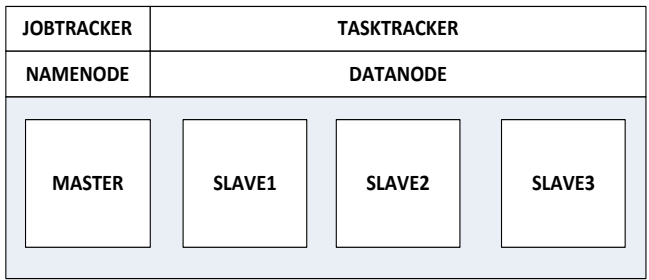

Fig. 3. Hadoop Cluster

## IV. GENETIC ALGORITHM

Genetic Algorithms (GAs) are a class of adaptive stochastic optimization algorithms involving randomness. The basic idea is to try to mimic natural selection [14-16]. The randomness is a controllable process. GAs have several advantages such as continuity and linearity. A genetic algorithm requires three basic processes: selection of parents, a mating between the parents and a fittest mechanism for survival. The main units are chromosome, gens, population, selection, crossover mutation and fitness function.

There are several coding techniques and they vary according to programmer and the problem. In this study, we used the following configurations and parameters for our GA implementation;

**Table I. The GA Configuration**

| Coding | Continuous GA |
|---|---|
| Mutation Rate | 1 % |
| Crossover Rate | 80 % |
| Crossing Method | Haupt's method |
| Selection Method | Rank List |
| Population | Between 1500 and 10 million |
| Iteration | 1000 |

## V. EXPERIMENTS

The first experiments were tried with a simple genetic algorithm using populations of different size for a sphere function (D=300), running on a single PC having 8 GB RAM and a 64 bit OS. As seen in Fig. 4 and Fig. 5, the PC threw out of memory exception for the population of 1 million. Then a sequential test was done. Bu this time we faced long processing times, for example 1 million second for the population of 1 million.

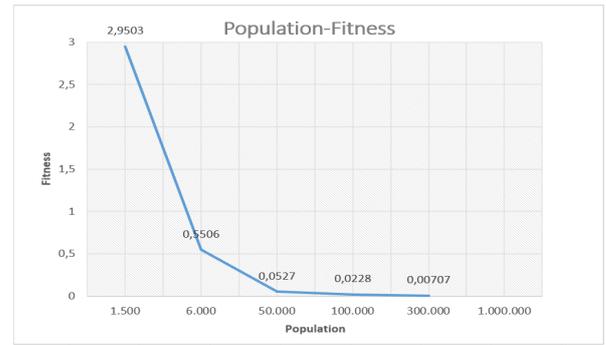

Fig. 4. The Populations and the results for single PC

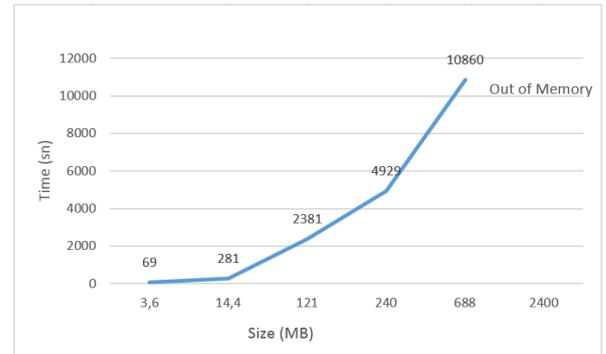

Fig. 5. Process Times according to the populations

Consecutively a MapReduce implementation of the same GA was run on the Hadoop cluster for various population sizes. In these experiments the system was able to operate successfully for each population. However, increase in the size of the population did not affect the results positively. In fact, this was an expected result, because of the fact that big populations were broken down into fixed-sized blocks and MapReduce jobs were run on these blocks, which in turn caused the results to remain almost same for all the iterations. Therefore we conclude that the block size is a factor which can directly affect the precision.

**Table II. The size and MEr Values of the populations**

| Population | Size (MB) | Result(MEr) | Time (sn) | Map |
|---|---|---|---|---|
| 1.500 | 3 | 3,1256 | 172 | 1 |
| 6.000 | 11 | 0,6101 | 403 | 1 |
| 80.000 | 138 | 0,0251 | 5920 | 1 |
| 140.000 | 240 | 0,0282 | 5520 | 2 |
| 500.000 | 860 | 0,0289 | 5898 | 7 |
| 1.000.000 | 1.700 | 0,0265 | 13762 | 14 |
| 5.000.000 | 8.500 | 0,0262 | 56748 | 68 |
| 10.000.000 | 17.000 | 0,0272 | 119983 | 135 |

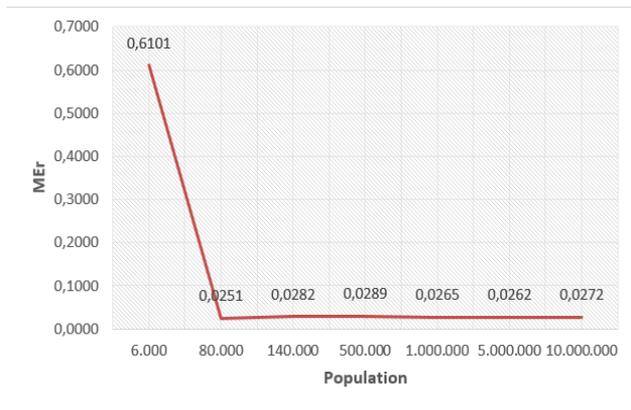

Fig. 6. The results of basic MRGA for different populations

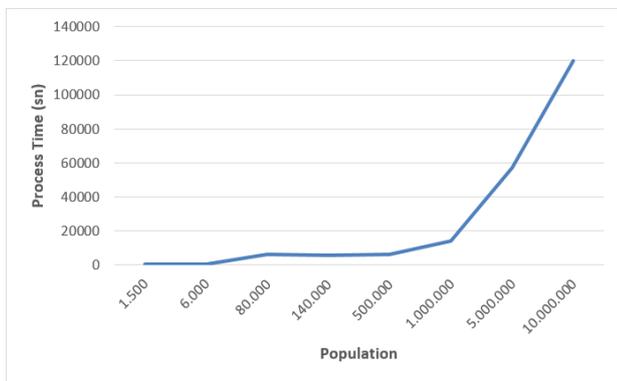

Fig. 7. Process Times of MRGA according to the populations

As seen in the Fig. 6, increase in the size of the population did not affect the solutions.

In the following experiments, we used a simple technique in which an extra GA run on the reduce phase. The population of the GA was obtained from the final populations in the maps by a small selection rate. The experiments were performed with populations of 1, 5 and 10 million. With these experiments, we achieved better results in line with the increase in the populations (Fig.8).

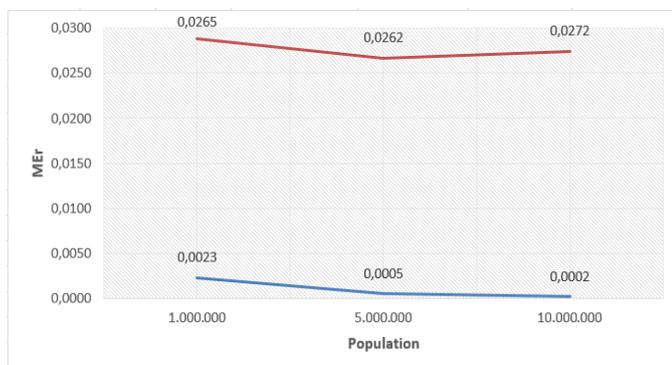

Fig. 8. The results of the proposed technique and the basic MRGA

## VI. CONCLUSION

In this study, we show that a simple genetic algorithm using very big populations, which run on a parallel and distributed system, can be used for solving high-dimensional problems. We used open source Hadoop platform and Sphere benchmark function. In the initial experiments, we saw that after a certain point the increase in the size of the population does not change the solution. With the help of a simple but effective approach we developed, some improvements were made. As seen in the results, the genetic algorithm running on the reduce phase, which used the selected and transferred individuals from the populations of the Maps by a low SR, had a high success rate and thus the increase in the size of the population affected the results positively.

However, in the end of our experiments we drew a conclusion that although better results could be obtained, a simple genetic algorithm using very big populations is not adequate to apply to high-dimensional problems in terms of process time and workload. The fact remains that it is still possible to obtain more successful results by employing different improvement techniques or hybrid methods. Specifically, use of MapReduce chains, the dynamic generation of populations in the map phase or the use of other improved genetic algorithms may be analyzed.